\documentclass[technote,10pt]{IEEEtran}
\usepackage{cite}
\usepackage[cmex10]{amsmath}
\usepackage{graphicx}
\usepackage{picinpar}
\usepackage{amsmath,amsfonts,amssymb}
\usepackage{array}
\usepackage{mdwmath}
\usepackage{mdwtab}
\usepackage{eqparbox}
\usepackage{stfloats}
\usepackage{subfig}
\usepackage{ragged2e}
\usepackage{extarrows}
\usepackage{booktabs}
\usepackage{multirow}
\usepackage{etoolbox}
\usepackage{makecell}
\usepackage{graphicx}
\usepackage{float}
\usepackage{subfig}
\usepackage{overpic}

\usepackage{algorithm}
\usepackage{algorithmic}
\usepackage{fixltx2e}
\usepackage{xcolor}
\usepackage{bm}
\usepackage{geometry}
\begin{document}
\newtheorem{lemma}{Lemma}
\newtheorem{corol}{Corollary}
\newtheorem{theorem}{Theorem}
\newtheorem{proposition}{Proposition}
\newtheorem{definition}{Definition}
\newcommand{\e}{\begin{equation}}
\newcommand{\ee}{\end{equation}}
\newcommand{\eqn}{\begin{eqnarray}}
\newcommand{\eeqn}{\end{eqnarray}}
\renewcommand{\algorithmicrequire}{ \textbf{Input:}} 
\renewcommand{\algorithmicensure}{ \textbf{Output:}} 

\newenvironment{shrinkeq}[1]
{ \bgroup
\addtolength\abovedisplayshortskip{#1}
\addtolength\abovedisplayskip{#1}
\addtolength\belowdisplayshortskip{#1}
\addtolength\belowdisplayskip{#1}}
{\egroup\ignorespacesafterend}
\title{MmWave-LoRadar Empowered Vehicular Integrated
Sensing and Communication Systems: LoRa Meets FMCW}
\author{Yi Tao, Ziwei Wan, Zhuoran Li, Zhen Gao, {\textit{Member, IEEE}}, and Gaojie Chen, {\textit{Senior Member, IEEE}}, Rui Na, {\textit{Student Member, IEEE}}
\vspace{-5mm}
\thanks{The work was supported in part by the Natural Science Foundation of China (NSFC) under Grant 62471036, Beijing Natural Science Foundation under Grant L242011, Shandong Province Natural Science Foundation under Grant ZR2022YQ62, and Beijing Nova Program. {\it{(Corresponding author: Zhen Gao; Rui Na.)}}}
\thanks{Yi Tao and Ziwei Wan contribute to this paper equally.}
\thanks{Yi Tao, Ziwei Wan, Zhuoran Li, and Zhen Gao are with the School of Information and Electronics, Beijing Institute of Technology, Beijing 100081, China (e-mail: \{taoyi2022, ziweiwan, lizhuoran23,  gaozhen16\}@bit.edu.cn).

Gaojie Chen is with the School of Flexible Electronics (SoFE) \& State Key Laboratory of Optoelectronic Materials and Technologies (OEMT), Sun Yat-sen University, Shenzhen 518107, China (e-mail: gaojie.chen@ieee.org).

Rui Na is with the Advanced Research Institute of Multidisciplinary Sciences, Beijing Institute of Technology, Beijing 100081, China, also with the Yangtze Delta Region Academy of Beijing Institute of Technology, Jiaxing 314000, China (e-mail: narui@bit.edu.cn).
	}
}
\maketitle

\begin{figure}[t]
	\captionsetup{font=footnotesize,  name = {Fig.}, labelsep = period}
	\centering	\includegraphics[width=6.5cm, keepaspectratio]%
	{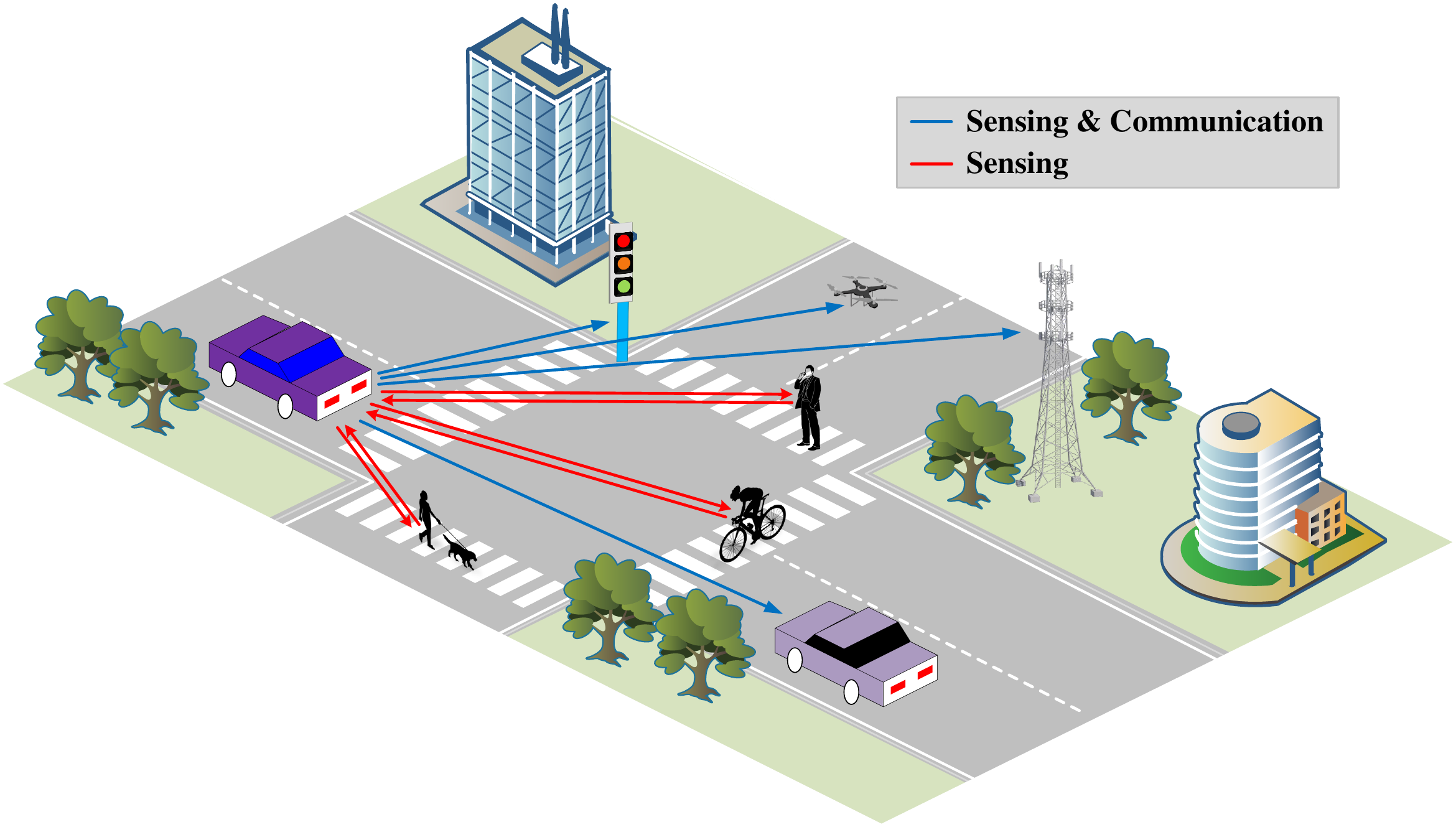}
	\caption{An illustration of the mmWave-LoRadar ISAC system for one car in vehicular applications.}
	\vspace*{-7mm}
\end{figure}
\begin{abstract}
The integrated sensing and communication (ISAC) technique is regarded as a key component in future vehicular applications. In this paper, we propose an ISAC solution that integrates Long Range (LoRa) modulation with frequency-modulated continuous wave (FMCW) radar in the millimeter-wave (mmWave) band, called mmWave-LoRadar. This design introduces the sensing capabilities to the LoRa communication with a simplified hardware architecture.
Particularly, we uncover the dual discontinuity issues in time and phase of the mmWave-LoRadar received signals, rendering conventional signal processing techniques ineffective. As a remedy, we propose a corresponding hardware design and signal processing schemes under the compressed sampling framework.
These techniques effectively cope with the dual discontinuity issues and mitigate the demands for high-sampling-rate analog-to-digital converters while achieving good performance. Simulation results demonstrate the superiority of the mmWave-LoRadar ISAC system in vehicular communication and sensing networks.
\end{abstract}
 \vspace*{-1mm}
 \vspace*{-2.0mm}
\IEEEpeerreviewmaketitle
\vspace*{-4mm}
\section{Introduction}
Radar sensing and wireless communication, as two pivotal technologies based on electromagnetic waves, share similarities in hardware structures and algorithmic processing \cite{xingyuzhou}. Consequently, the concept of integrated sensing and communication (ISAC) has been proposed and identified as a key technology in the sixth-generation (6G) networks \cite{zyf,lhs,mao1,mao2}. Particularly for future Internet of Vehicles (IoV), it is essential to ensure data transmission efficiency and high-precision radar sensing under high-mobility scenarios. 

Among several ISAC waveform design approaches, the sensing-centric waveform design has been proven to be highly adaptable for vehicular communication and sensing \cite{JRC}, offering notable advantages in strong resistance to interference, low-cost implementation, and low latency. Currently, numerous sensing-centric waveforms have been proposed to realize communication functions utilizing the frequency-modulated continuous wave (FMCW). 
In \cite{FRaC}, a data transmission method was proposed that combines carrier selection, antenna selection, and phase modulation with FMCW. However, the modulation and demodulation process for the antenna selection is of high complexity. Furthermore, the utilization of time-frequency resources also requires a thorough investigation.

In addition to FMCW radar, chirp techniques have been introduced to communication waveform design. One typical example is the Long Range (LoRa) modulation \cite{LoRa}. Due to the adoption of chirp waveform, LoRa inherently possesses sensing capabilities and thus is regarded as a promising waveform for ISAC. For instance, in \cite{LoRadar}, the proposed LoRadar system generated two FMCWs with a time offset, producing the harmonic signal that serves as the carrier frequency of LoRa. However, the harmonic signal is usually weaker than the transmit signal, which significantly reduces the maximum communication range. Furthermore, with sensing and communication being implemented in separate frequency bands, achieving a fully integrated waveform design poses an important challenge. Considering spectrum sharing, note that vehicular FMCW radar works at the millimeter-wave (mmWave) frequency band. Whereas, if working at the mmWave band with a large bandwidth, conventional LoRa processing would incur prohibitive hardware costs. Therefore, integrating LoRa modulation into mmWave FMCW radar needs to be well investigated.
In this paper, we propose a novel ISAC waveform by integrating LoRa with mmWave FMCW, termed mmWave-LoRadar.
Due to data modulation solely on the starting frequency, the scheme demonstrates exceptional scalability and can be further integrated with other existing schemes.
In Fig. 1, one typical mmWave-LoRadar ISAC system is depicted.
Specifically, our contributions are summarized as follows.

\begin{itemize}

    \item{For sensing tasks, we propose a pseudo-random time-division multiplexing (TDM) multiple-input multiple-output (MIMO) transmission scheme. The {\it dual discontinuity} of the intermediate frequency (IF) signal in terms of both time and phases is revealed. 
    To save hardware cost and energy consumption, we utilize analog-to-information converters (AICs) to sample the IF signals.}
    Furthermore, we propose a phase compensation (PC)-based compressed sampling (CS) velocity estimation to prevent the degradation of velocity resolution and velocity ambiguity and eliminate phase errors induced by target motion when estimating angles.
    
 
    \item{For communication tasks, the communication ability of LoRa is well reserved by embedding data into the spectrum-folded chirp signals. Considering the demand for large bandwidth sampling and the sparsity of signals, we also employ AICs and propose a CS-based data demodulation scheme, which effectively reduces the hardware burden of the mmWave-LoRadar system while enhancing system efficiency.
    Additionally, note that a lot of hardware (including AICs) can be shared between sensing and communication, indicating the proposed system to be a promising ISAC solution.

    }

\end{itemize}
 
\emph{Notations:} Column vectors and matrices are denoted by boldface lower and upper-case symbols, respectively. $\mathbb{C}$ is the set of complex numbers.
$\varnothing$ is the empty set. ${\rm{min}}\{\cdot\}$ and ${\rm{max}}\{\cdot\}$ are the minimum and maximum values, respectively.
$\left\lfloor\cdot\right\rfloor$ is the flooring function. ${\left\langle \cdot \right\rangle _N}$ denotes the remainder after divided by $N$. $\odot$ is the Hadamard product. 
$\left[{\bf{A}}\right]_{\mathcal{I},:}$ ($\left[{\bf{A}}\right]_{:,\mathcal{I}}$) denotes the sub-matrix consisting of the rows (columns) of ${\bf{A}}$ indexed by the set $\mathcal{I}$.
$\left[{\bf{a}}\right]_{\mathcal{I}}$ denotes the vector constructed from the elements of ${\bf{a}}$ indexed by $\mathcal{I}$.
${\bf{I}}_N$ is the $N$-order identity matrix.
$(\cdot)^{-1}$, $(\cdot)^{\rm{T}}$, and $(\cdot)^{\rm *}$ are the inverse, transpose, and conjugate operations, respectively.
 \vspace*{-3.0mm}
\section{System Model}
The mmWave-LoRadar ISAC system consists of an ISAC transmitter (ISAC-Tx) with $L_{t}$ transmit antennas (Tx-As), a sensing receiver (Sen-Rx), and a communication receiver (Com-Rx) that share $ L_{r} $ receive antennas (Rx-As), as shown in Fig. 2. 
We consider the uniform linear array (ULA) for both transmit and receive arrays.
The antenna spacing of the receive array is $ d = \lambda/2 $ while that of ISAC-Tx is $ L_{r}d $, where $ \lambda $ is the wavelength corresponding to the carrier frequency $ f_{c} $. 
This configuration is intended to form a virtual ULA with $L_t L_r$ antennas and spacing $d$ \cite{vr}.

Fig. 2 illustrates the hardware structure of the mmWave-LoRadar.
During the ISAC process, the ISAC-Tx generates symbols at the mmWave-LoRadar symbol generator and transmits them via a single Tx-A selected by the antenna switch.
The sensing terminal receives echo signals and feeds them back to the Sen-Rx for sensing processing, and the communication terminal receives signals and uses the Com-Rx for communication processing.
Benefiting from the shared hardware, the system can switch between different tasks via function switches.
Furthermore, by maintaining consistency with TDM-MIMO FMCW radar, the mmWave-LoRadar exhibits characteristics of low hardware cost and low power consumption, and it does not require the addition of any expensive extra components.
Especially during the sampling process, although AIC has a much lower sampling rate than ADC, parameter estimation and data demodulation can still be achieved via CS algorithms, thereby greatly reducing hardware costs and power consumption.


A holistic waveform based on LoRa is utilized for the proposed mmWave-LoRadar. Assuming that $P$ symbols are transmitted in one frame, with each symbol comprising a payload part of length $T$ and a guard interval (GI) of length $T_{\rm GI}$, resulting in the symbol duration $T_{0} = T + T_{\rm GI}$. 
The spreading factor $N_{\rm SF}$ is employed to define $H = 2^{N_{\rm{SF}}}$ different frequency shifts \cite{LoRa}, embedding $N_{\rm SF}$-bit data into each symbol. Additionally, the bandwidth $B$ must satisfy $B=H/T$ to guarantee the quasi-orthogonality among symbols with different frequency shifts. Therefore, the $p$th transmit symbol, $p \in \{ 0,1,...,P-1\}$, is represented as
\begin{align}
\label{equ:sp}
    \hspace{-1mm} {s_p}\left( t \right) = {\rm{rect}}\left( {\frac{{t - p{T_0}}}{T}} \right){e^{j2\pi \left[ {{f_c}\left( {t - p{T_0}} \right) + \int_0^{t - p{T_0}} {{\phi _p}\left( x \right){\rm{d}}x} } \right]}},
\end{align}
where ${\rm{rect}} \left( t \right) \buildrel \Delta \over =  1 $ for $t \in \left[ 0,1 \right)$ and $0$ otherwise, ${\phi _p}\left( x \right) = {\left\langle {\frac{B}{T}x + \frac{{{h_p}}}{T}} \right\rangle _B} - \frac{B}{2}$, $h_p \in \left\{ 0,1,...,H-1 \right\}$ represents the index of frequency shift used by the $p$th symbol, indicating the data carried by the $p$th symbol. 
\begin{figure}[t]
	\captionsetup{font=footnotesize, name = {Fig.}, labelsep = period}
	\centering
	{\includegraphics[width=7.5cm, keepaspectratio]
		{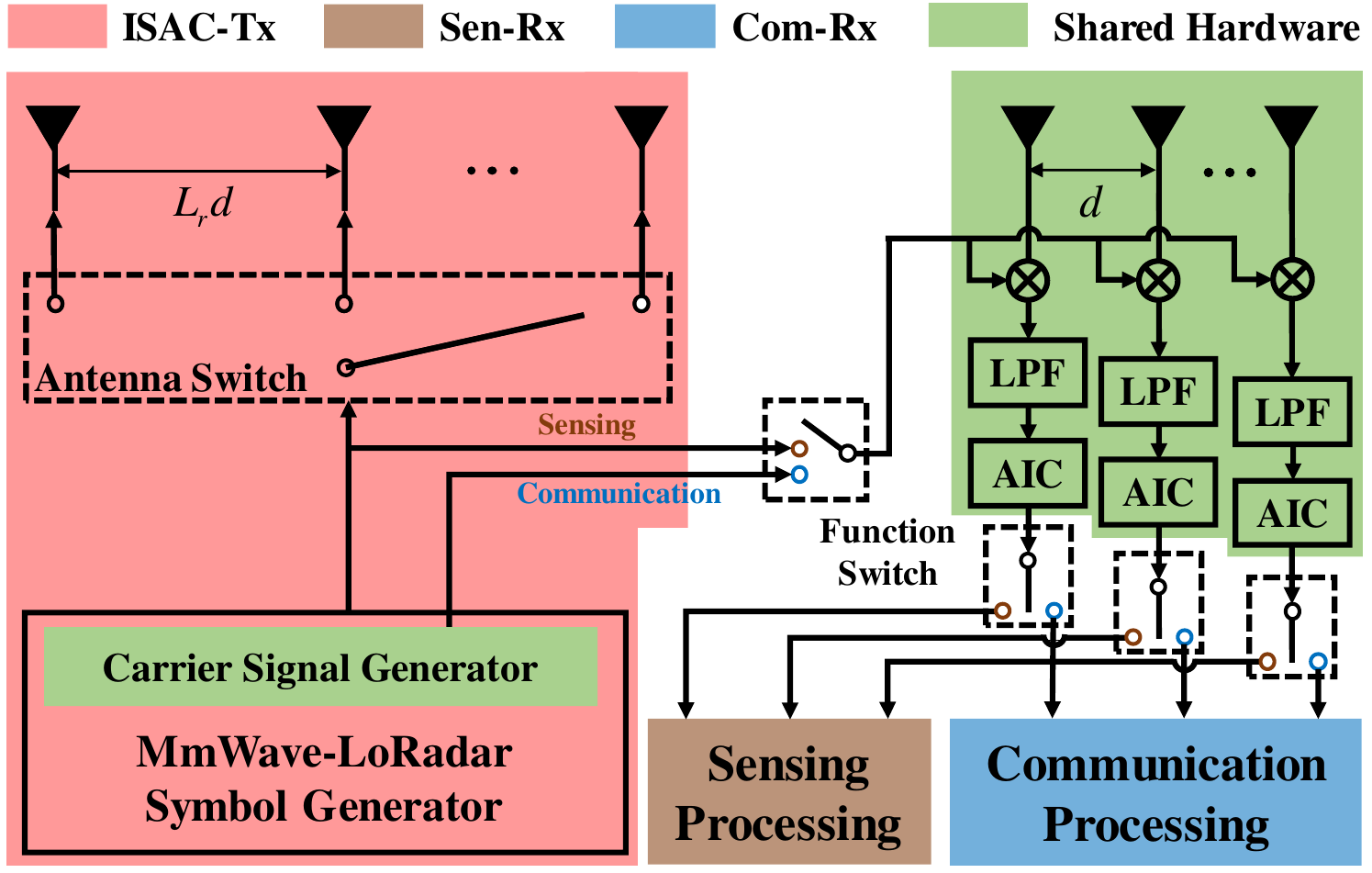}}	
 \caption{The hardware architecture of the mmWave-LoRadar.}
 \vspace{-6mm}
\end{figure}

Inspired by \cite{GZ, Random}, to harness the superiority of CS in TDM-MIMO, a pseudo-random TDM-MIMO transmission scheme is considered.
As exemplified in Fig. 3, unlike the conventional TDM-MIMO scheme \cite{v&r}, Tx-As for each transmit symbol are pseudo-randomly selected to guarantee the effectiveness of the subsequent CS processing.
We denote $l_p \in \left\{ 0,1,...,L_{ t}-1 \right\}$ as the index of the activated Tx-A for the $p$th symbol.
For sensing applications, we assume $K$ point targets. Given sufficiently long $T_{\rm{GI}}$, the ISI-free echo signal for the $p$th symbol ${{\bf{r}}_p}\left( t \right) \in \mathbb{C}^{L_{ r} }$ at the Rx-A array can be expressed as
\begin{align}
\label{equ:rp}
    {{\bf{r}}_p}\left( t \right) = \sum\limits_{k = 1}^K {{\alpha _k}{\bf{a}}\left( {{\theta _k}} \right){s_p}\left( {t - {\tau _k} - \frac{2v_k}{c}t} \right){e^{j\pi  {{l_p} } {L_{ r}}\sin {\theta _k}}}} ,
\end{align}
where $c$ is the speed of light, $\alpha_k$, $\theta_k$, $\tau_k$, and $v_k$ represent the gain, angle, delay, and relative radial velocity, respectively, of the $k$th target, and ${\bf{a}}\left( \theta  \right) = {\left[ {1,{e^{j\pi \sin \theta }},...,{e^{j\pi \left( {{L_{ r}} - 1} \right)\sin \theta }}} \right]^{\rm{T}}} \in \mathbb{C}^{L_{ r} }$ is the steering vector of the Rx-A array. Note that $\tau_k \buildrel \Delta \over = 2r_k/c$ with $r_k$ being the range of the $k$th target and ${T_{{\rm{GI}}}} \gg \mathop {\max }\limits_k \left( {{\tau _k}} \right)$. For brevity, we omit the noise factor in \eqref{equ:rp}.

For communication, signals from another device's ISAC-Tx are received by Rx-As and further employed for communication tasks, e.g., channel estimation and data demodulation. For simplicity, we focus on the process in a single Rx-A.
Given the high attenuation in mmWave channels, only the line of sight (LOS) communication channel is considered. Therefore, similarly as \eqref{equ:rp}, the received signal for the $p$th communication symbol ${\bar r_p}\left( t \right)$ can be written as ${\bar r_p}\left( t \right) = {\bar \alpha}{s_p}\left( {t - \bar \tau - {\tau_{\rm{syn}}} - \bar v t /c } \right)$, where ${\tau_{\rm{syn}}}$ is the synchronization delay, $\bar {\alpha }$, ${\bar \tau}$, and $\bar v$ are the gain, delay, and relative radial velocity of the LOS component, respectively.
\vspace{-5mm}
\section{Sensing Processing of MmWave-LoRadar}

\subsection{IF Signal after Mixing and CS AIC at Sen-Rx}
For the sensing processing, we refer to FMCW radar technology \cite{automotive} in mmWave-LoRadar. 
Particularly, the echo signal ${{\bf{r}}_p}\left( t \right)\in \mathbb{C}^{L_{ r} }$ is conjugately mixed with the corresponding transmit signal via mixers to generate the IF signal. Note that for the $p$th symbol, only the IF signal within $t \in \left[ {p{T_0} + {T_{{\rm{mix}}}},p{T_0} + T} \right)$ is considered to avoid the vacancy caused by the delay, where ${T_{{\rm{mix}}}} > \mathop {\max }\limits_k \left( {{\tau _k}} \right)$. 
Unlike the conventional FMCW radar with $h_p = 0$, $\forall p$, the mmWave-LoRadar symbol may exhibit unique properties at the time-frequency plane, as depicted in Fig. 3. During mixing, the frequency of the corresponding IF signal may be much higher than anticipated. 
Specifically, the IF signal ${{\bf{r}}_p^{{\rm{IF}}}\left( t \right)} \in \mathbb{C}^{L_{ r} }$ can be expressed as
\begin{align}
\label{equ:sIFp}
    {\bf{r}}_p^{{\rm{IF}}}\left( t \right) & = {{\bf{r}}_p}\left( t \right)s_p^ * \left( t \right) \nonumber \\
    & = \sum\limits_{k = 1}^K {\alpha' _k}{\bf{a}}\left( {{\theta _k}} \right){e^{j\pi {{l_p}} {L_{{r}}}\sin {\theta _k}}}  {e^{j2\pi {\mu_k}t}}  {{r}}_{p,k}^{{\rm{IF}}}\left( t \right),
\end{align}
where ${\alpha' _k} \buildrel \Delta \over = {\alpha _k}{e^{ - j2\pi {f_c}\tau_k}}$, $\mu_k \buildrel \Delta \over = 2v_k/ \lambda$ is the Doppler frequency of the $k$th target, and ${{r}}_{p,k}^{{\rm{IF}}}\left( t \right)$ is defined as
\begin{equation}
\label{equ:sIFpk}
\begin{aligned}
r_{p,k}^{{\rm{IF}}}\left( t \right) 
&{\overset{1}{\operatorname*{\approx}} } \left\{ {\begin{array}{*{20}{l}}
{{e^{j{\varphi _{p,k}}}}{e^{ - j{2\pi f_k^{{\rm{IF}}}\left(t-pT_0\right)} }},}&{t \in {{\Omega }}_{p,k}^{\rm{t_1}}}\\
{{e^{j{\varphi_{p,k}}}}{e^{ - j{2\pi \left(f_k^{{\rm{IF}}}-B\right)\left(t-pT_0\right)} }},}&{t \in {{\Omega}}_{p,k}^{\rm{t_2}}} \\
{{e^{j\left({\varphi_{p,k}}+2\pi B\tau_k\right)}}{e^{ - j{2\pi f_k^{{\rm{IF}}}\left(t-pT_0\right)} }},}&{t \in {{\Omega}}_{p,k}^{\rm{t_3}}}
\end{array}} \right. ,
\end{aligned}
\end{equation}
where ${\overset{1}{\operatorname*{\approx}} }$ is because the term $\frac{2v_k}{c}t$ is usually much smaller than $\tau_k$ and thus discarded \cite{v&r}, ${{f_k^{{\rm{IF}}}}} \buildrel \Delta \over = {{{B{\tau _k}}}/{T}}$ is the IF component, and the sets ${{\Omega}}_{p,k}^{\rm{t_1}}$, ${{\Omega}}_{p,k}^{\rm{t_2}}$, and ${{\Omega}}_{p,k}^{\rm{t_3}}$ are defined as $\left[ {p{T_0} + {T_{{\rm{mix}}}},p{T_0} + T_{p,k}^{{\rm{BWS}}}} \right)$, $\left[ {p{T_0} + T_{p,k}^{{\rm{BWS}}},p{T_0} + T_{p,k}^{{\rm{BWE}}}} \right)$, $\left[ {p{T_0} + T_{p,k}^{{\rm{BWE}}},p{T_0} + T} \right)$, respectively, while
\begin{align}
\label{equ:Tpk}
    T_{p,k}^{{\rm{BWS}}} & = \left\{ {\begin{array}{*{20}{l}}
{{T_{{\rm{mix}}}},}&{{\tau _k} > {{{T - {h_p}/B}}}}\\
{{{ {T - {h_p}/B} }},}&{{\text{otherwise}}}
\end{array}} \right.,\\
    T_{p,k}^{{\rm{BWE}}} & = \min \left\{ {T,\max{ \left\{ {\tau _k} + {T-{h_p}/B},{T_{\rm{mix}}}\right\}}}  \right\},
\end{align}
denote the blank window start (BWS) time and blank window end (BWE) time, respectively, and
\begin{align}
\label{equ:phi0}
{{{\varphi _{p,k}}}} & =  \pi\left({ {{f_k^{{\rm{IF}}}}}} + B -2 h_p/T \right) \tau_k.
\end{align}
We can observe that different values of $h_p$ and $\tau_k$ result in different $T_{p,k}^{{\rm{BWS}}}$ and $T_{p,k}^{{\rm{BWE}}}$.
Based on Fig. 3, we provide a detailed explanation of the behavior of the IF signal. We select $p=1$, $p=2$, and $p=3$ as examples, representing cases where only the second chirp segment is effective, both are effective, and only the first one is effective, respectively.
Three typical cases are illustrated as follows.
\begin{itemize}
	\item{
When $p=1$, $T_{{\rm{mix}}} \geq {T - {h_1}/B}$. In this case, if ${\tau _k} > {T - {h_1}/B}$, then blank window is $T_{1,k}^{{\rm{BWS}}}={T_{{\rm{mix}}}}$ and $T_{1,k}^{{\rm{BWE}}}=\max{ \left\{ {\tau _k} + {T-{h_1}/B},{T_{\rm{mix}}}\right\}}$. If ${{\tau _k} \leq {{T - {h_1}/B}}}$, $T_{1,k}^{{\rm{BWS}}}={{T - {h_1}/B}}$ and $T_{1,k}^{{\rm{BWE}}}=\max{ \left\{ {\tau _k} + {T-{h_1}/B},{T_{\rm{mix}}}\right\}}$. If $T_{{\rm{mix}}} \geq  {T - {h_1}/B} + {\tau _k}$, for these two situations, ${{\Omega}}_{1,k}^{\rm{t_1}} = \varnothing$ and ${{\Omega}}_{1,k}^{\rm{t_3}} \neq \varnothing$. ${{\Omega}}_{1,k}^{\rm{t_2}} = \varnothing$ for ${\tau _k} \geq {T - {h_1}/B}$ and ${{\Omega}}_{1,k}^{\rm{t_2}} \neq \varnothing$ for ${\tau _k} \geq {T - {h_1}/B}$, respectively. Therefore, effective sampling starts from $p{T_0}+T_{{\rm{mix}}}$.
If $T_{{\rm{mix}}} <  {T - {h_1}/B} + {\tau _k}$, ${{\Omega}}_{1,k}^{\rm{t_1}} = \varnothing$, ${{\Omega}}_{1,k}^{\rm{t_2}} \neq \varnothing$, and ${{\Omega}}_{1,k}^{\rm{t_3}} \neq \varnothing$. Therefore, effective sampling starts from $p{T_0}+{T - {h_1}/B} + {\tau _k}$. 
}
\item{
When $p=2$, $T_{\rm{mix}}<{T - {h_2}/B}$ and ${h_2}/B \geq {\tau _k}$.
$T_{2,k}^{{\rm{BWS}}}={T - {h_2}/B}$ and $T_{2,k}^{{\rm{BWE}}}={\tau _k} + {T - {h_2}/B}$.
Therefore, ${{\Omega}}_{2,k}^{\rm{t_1}} \neq \varnothing$, ${{\Omega}}_{2,k}^{\rm{t_2}} \neq \varnothing$, and ${{\Omega}}_{2,k}^{\rm{t_3}} \neq \varnothing$. }
\item{
Finally, $p=3$, $T_{\rm{mix}}<{T - {h_3}/B}$ and ${h_3}/B < {\tau _k}$. Consequently, $T_{3,k}^{{\rm{BWS}}}={T - {h_3}/B}$ and $T_{3,k}^{{\rm{BWE}}}=T$.
Therefore, ${{\Omega}}_{3,k}^{\rm{t_1}} \neq \varnothing$, ${{\Omega}}_{3,k}^{\rm{t_2}} \neq \varnothing$, and ${{\Omega}}_{3,k}^{\rm{t_3}} = \varnothing$.
}
\end{itemize}
Note that the gray area in Fig. 3 represents the effective sampling time intervals that correspond to the non-empty parts within ${{\Omega}}_{p,k}^{\rm{t_1}}$, ${{\Omega}}_{p,k}^{\rm{t_2}}$, and ${{\Omega}}_{p,k}^{\rm{t_3}}$.
\begin{figure}[t]
	\captionsetup{font=footnotesize, name = {Fig.}, labelsep = period}
	\centering
	{\includegraphics[width=8.4cm, keepaspectratio]
		{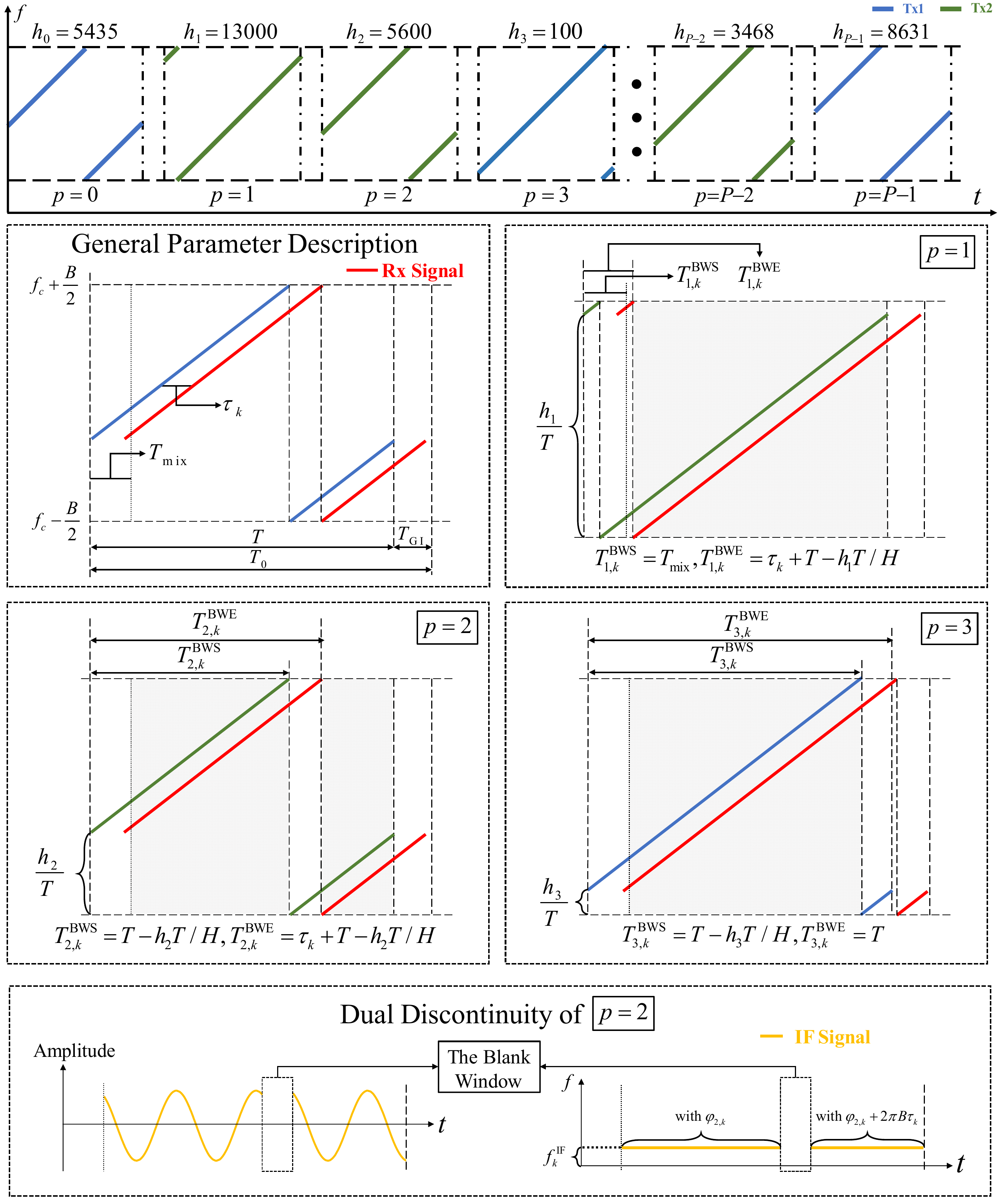}}	
	\caption{An example of the mmWave-LoRadar utilizing $2$ Tx-As, where $H = 2^{14}$, $B=1$\,GHz, $T=16.4\ \mu$s, and $T_{\rm{GI}} = T_{\rm{mix}}=0.5\ \mu $s.}
	\vspace*{-5mm}
\end{figure}

The output of each mixer is then fed into an analog low-pass filter (LPF) whose cutoff frequency is set to $f_{\max} \ge B\mathop {\max }\limits_k \left( {{\tau _k}} \right)/T $. Consequently, the IF component with the higher frequency ${f_k^{{\rm{IF}}}-B}$ is filtered out, whereas that with the frequency $f_k^{{\rm{IF}}}$ is preserved. Considering the ideal analog LPF, the output corresponding to \eqref{equ:sIFp} can be written as
\begin{align}
\label{equ:sLPFp}
    {\bf{r}}_p^{{\rm{LPF}}}\left( t \right) = \sum\limits_{k = 1}^K {{{\alpha_{k}'}}{\bf{a}}\left( {{\theta _k}} \right){e^{j\pi {l_p}{L_{{r}}}\sin {\theta _k}}}{e^{j2\pi {\mu_k}t}}r_{p,k}^{{\rm{LPF}}}\left( t \right)} ,
\end{align}
where $r_{p,k}^{{\rm LPF}}\left( t \right)$ is given as
\begin{align}
\label{equ:sLPFpk}
r_{p,k}^{{\rm{LPF}}}\left( t \right) =
\left\{ {\begin{array}{*{20}{l}}
{0,}&{t \in {{\Omega}}_{p,k}^{\rm{t_2}}} \\
{{r_{p,k}^{\rm IF}\left( t \right)},}&{\text{otherwise}}
\end{array}} \right. .
\end{align}

Obviously, the IF signals of mmWave-LoRadar may exhibit dual discontinuity, which is distinctive compared to the conventional FMCW radar system with $h_p = 0$. Specifically, the dual discontinuity consists of 1) {\bf time discontinuity}, introducing the blank window (all-zero segment) to the IF signals, and 2) {\bf phase discontinuity}, i.e., there is a phase shift of $2\pi B \tau_k$ between different segments of the IF signals, as shown in \eqref{equ:sIFpk}. This unique property causes model mismatch and thus invalidates many off-the-shelf sensing algorithms. Fortunately, the inherent frequency sparsity of mmWave sensing channels can be preserved under the dual discontinuity, which motivates the use of CS techniques.
In this paper, we consider the utilization of AIC \cite{AIC}. 
AIC is a low hardware cost and low power consumption sampling device based on the theory of CS, which significantly reduces the sampling rate while maintaining performance in high bandwidth frequency ranges.
Additionally, AICs can be shared between communication and sensing, as discussed in Section {\rm{I}} and {\rm{II}}\footnote{Several studies have proposed transistor-level implementation schemes and experimentally validated the applications of AIC. Nevertheless, the focus of this paper is not on hardware design and implementation but rather on the theoretical analysis of the AIC. Hence, we do not elaborate on the design details of AIC but rather focus on the CS applications.}.
We denote $N$ as the number of sampling points for each symbol, and ${{\cal M}} = \left\{ {{m_{0}},{m_{1}},...,{m_{N-1}}} \right\}$ as the random sampling index set of AIC, where $0 \le {{m_{0}}<{m_{1}}<...<{m_{N-1}}} \le N_{\max}-1$ and ${N_{\max }} \buildrel \Delta \over = \left\lfloor {f_{\max}}{\left( {T - {T_{{\rm{mix}}}}} \right)} \right\rfloor $. The signal of the $n$th sample in the $p$th symbol, denoted as ${\bf{r}}_{p}^{{}}\left[ n \right] \in \mathbb{C}^{L_r}$, $n \in \left\{ {0,1,...,N - 1} \right\}$, can be represented as
\begin{align}
\label{equ:rADCpn}
{\bf{r}}_{p}^{{}}\left[ n \right] & = {\bf{r}}_p^{{\rm{LPF}}}\left( t \right)\left| {_{t = p{T_0} + {T_{{\rm{mix}}}} + {m_{n}}/{f_{{\max}}}}} \right. \nonumber \\
& = \sum\limits_{k = 1}^K {{{\alpha_{k}''}}{\bf{a}}\left( {{\theta _k}} \right){e^{j\pi {l_p}{L_{\rm{r}}}\sin {\theta _k}}}{e^{j2\pi p{\mu_k}{T_0}}} {r_{p,k}^{{}}\left[ n \right]}} .
\end{align}
where ${{\alpha_{k}''}} \buildrel \Delta \over = {\alpha _{k}'}{e^{j2\pi \left( {{\mu_k} - f_k^{{\rm{IF}}}} \right){T_{{\rm{mix}}}}}}$, and
\begin{equation}
\label{equ:sADCpkn}
   r_{p,k}^{{}}\left[ n \right]  {\overset{2}{\operatorname*{\approx}} }
  {e^{ - j {2\pi \frac{{f_k^{{\rm{IF}}}}}{{{f_{\max }}}}{m_{n}} } }} 
  \times \left\{ {\begin{array}{*{20}{l}}
{{e^{ j {{\varphi _{p,k}}} }},}&{{m_{n}} \in {{\Omega}}_{p,k}^{\rm{samp_1}}}\\
{0,}&{{m_{n}} \in {{\Omega}}_{p,k}^{\rm{samp_2}}} \\ 
{{e^{ j \left( {\varphi_{p,k}} + 2\pi B\tau_k \right) }},}&{{m_{n}} \in {{\Omega}}_{p,k}^{\rm{samp_3}}}
\end{array}}\right..
\end{equation}
In \eqref{equ:sADCpkn}, ${\overset{2}{\operatorname*{\approx}} }$ is because the Doppler component $\mu_k$ is much smaller than $f_k^{\rm IF}$ \cite{v&r}, and thus is discarded from $r_{p,k}^{{}}\left[ n \right]$. The sets ${{\Omega}}_{p,k}^{\rm{samp_1}}$, ${{\Omega}}_{p,k}^{\rm{samp_2}}$, and ${{\Omega}}_{p,k}^{\rm{samp_3}}$ are respectively defined as $\left[ 0, N_{p,k}^{{\rm{BWS}}} \right)$, $\left[ N_{p,k}^{{\rm{BWS}}},N_{p,k}^{{\rm{BWE}}} \right)$, and $\left[ N_{p,k}^{{\rm{BWE}}},N_{{\max}} \right)$, where $N_{p,k}^{{\rm{BWS}}}\buildrel \Delta \over=\left\lfloor{f_{\max }}\left( T_{p,k}^{{\rm{BWS}}} - {T_{{\rm{mix}}}} \right)\right\rfloor$ and $N_{p,k}^{{\rm{BWE}}}\buildrel \Delta \over=\left\lfloor{f_{\max }}\left( T_{p,k}^{{\rm{BWE}}} - {T_{{\rm{mix}}}} \right)\right\rfloor$. Note that the three sets in \eqref{equ:sADCpkn} correspond to the three sets in \eqref{equ:sIFpk}.

\vspace{-3mm}
\subsection{Range Estimation Using CS Sparse Signal Recovery}

The task of sensing would be range estimation (RE), velocity estimation (VE), and angle estimation (AE), respectively corresponding to estimating $\tau_k$, $v_k$, and $\theta_k$ from the compressed observations ${\bf r}_p \left[ n \right]$. 
We first focus on the RE. To this end, observations are collected to form the RE matrix ${{\bf{Y}}}_{\rm{RE}}=\left[{{\bf{R}}_0},{{\bf{R}}_1},...{{\bf{R}}_{P-1}}\right] \in \mathbb{C}^{N\times P L_r}$, where ${{\bf{R}}_p} = {\left[ {{\bf{r}}_p^{}\left[ 0 \right],{\bf{r}}_p^{}\left[ 1 \right],...,{\bf{r}}_{p}^{}\left[ N-1 \right]} \right]^{\rm{T}}} \in \mathbb{C}^{N\times L_r}$. Columns of ${\bf Y}_{\rm RE}$ are compressively sampled measurements of $K$ IF components with dual discontinuity. Resorting to the CS processing framework, ${\bf Y}_{\rm RE}$ can be modeled as
\begin{align}
\label{equ:Rcs}
{{\bf{Y}}}_{\rm{RE}} = {\bf{\Phi}} {\bf{\tilde F}}_{\rm{RE}} {\bf X},
\end{align}
where ${\bm{\Phi}} = {\left[{\bf{I}}_{N_{\max}}\right]}_{{\cal M,:}}\in \mathbb{C}^{N \times N_{\max}}$ is the measurement matrix, ${\bf{\tilde F}}_{\rm{RE}} \in \mathbb{C}^{{N_{\max}} \times \rho_{\rm RE} {N_{\max}}}$ with $\rho_{\rm RE} \ge 1$ is the dictionary matrix consisting of the first $N_{\max}$ rows of the $\left(\rho_{\rm RE} N_{\max}\right)$-order discrete Fourier transform (DFT) matrix,
and ${\bf X} \in {\mathbb{C}}^{{\rho_{\rm RE}} {N_{\max}} \times P L_{{r}}}$ is the sparse spectral components of IF signals. 
In the Fourier transform domain, each column vector of ${\bf X} $ has only a few non-zero elements, representing the range information of targets. In the process of CS, by solving an optimization problem, it is possible to recover ${\bf X} $ from the limited measurement ${{\bf{Y}}}_{\rm{RE}}$ to obtain RE.
Note that all the column vectors of ${\bf X}$ share the same sparse pattern, therefore the RE can be formulated as a multiple measurement vector (MMV) CS problem. MMV-OMP \cite{MMV-OMP} is employed to obtain the sparse estimate $\bf {\hat X}$. Subsequently, the constant false alarm rate (CFAR) detection is utilized by incorporating all the column vectors of $\bf {\hat X}$ to determine $K$ most significant rows. ${\hat n}_{k} \in \{0,1,...,\rho_{\rm RE}N_{\max}-1\}$ represents the index of the non-zero values corresponding to the $k$th target, therefore the delay estimate $\hat{\tau}_k$ of $\tau_k$ can be obtained as ${\hat n}_{k} f_{\max} T/\left( B\rho_{\rm RE}N_{\max} \right)$.
\vspace{-3mm}

\subsection{Velocity Estimation Using CS Sparse Signal Recovery}
In the conventional TDM-MIMO scheme, the time-division transmission of Tx-As causes velocity ambiguity, and the maximum unambiguous velocity is reduced by $L_t$ \cite{v&r}. In this paper, by combining pseudo-random antenna selection with CS algorithms, we are able to reconstruct the Doppler spectrum through sparse recovery, thereby avoiding the degradation of velocity resolution and maximum unambiguous velocity. Furthermore, this design can eliminate phase errors induced by target motion, therefore enhancing the accuracy of AE.
To perform VE, we construct the VE measurements associated with the $k$th target as ${\bf{\hat x}}_{k,r}^{\rm{T}} = {\left[ {{\bf{\hat X}}} \right]_{{{\hat n}_k},{{\cal I}_r}}}$ based on the $r$th Rx-A, $\bf{\hat X}$, and ${\hat n}_k$, where ${\bf{\hat x}}_{k,r} \in \mathbb{C}^P$, $r \in \{0,1,..., L_r-1\}$, and the ordered set ${\cal I}_r$ corresponding to the $r$th Rx-A is defined as ${\cal I}_r = \left\{ {r,r + {L_r},...,r + \left( {P - 1} \right){L_r}} \right\}$. 
For the accurate VE under phase discontinuity of the proposed mmWave-LoRadar, we propose the PC scheme based on the delay estimate of each target (referred to as PC $\rm{I}$).
The PC matrix ${\bf{R}}^{\rm{PC}}_{k} \in \mathbb{C}^{N\times P }$ for the $k$th target is defined as
\begin{align}
     {\left[ {{\bf{R}}_k^{{\rm{PC}}}} \right]_{n,p}} & = \left\{ 
     {\begin{array}{*{20}{l}}
 {e^{-j\hat{\varphi}_{p,k}},}&{{m_{n}} \in {{\Omega}}_{p,k}^{\rm{samp_1}}}\\
{1,}&{{m_{n}} \in {{\Omega}}_{p,k}^{\rm{samp_2}}}\\
{e^{-j\left(\hat{\varphi}_{p,k}+2\pi B \hat \tau_k\right)},}&{{m_{n}} \in {{\Omega}}_{p,k}^{\rm{samp_3}}}\\
 \end{array}} , \right.
 \end{align}
where $\hat{\varphi}_{p,k}$ is obtained by replacing $\tau_k$ with ${\hat \tau}_k$ in \eqref{equ:phi0}.
The VE measurements after PC are
\begin{align}
    {\bf{\hat x}}_{k,r}^{{\rm{PC}}} = {\bf{\hat x}}_{k,r}^{} \odot \left[ {{{\left( {{\bf{R}}_k^{{\rm{PC}}}} \right)}^{\rm{T}}}{{\bf{b}}_k}} \right] \in \mathbb{C}^{P},
\end{align}
where ${{\bf{b}}_k} = {\left[ {{\bf{\Phi \tilde F}}_{{\rm{RE}}}^*} \right]_{:,{{\hat n}_k}}} \in \mathbb{C}^{P}$.
Furthermore, we re-arrange the elements in ${\bf{\hat x}}_{k,r}^{{\rm{PC}}}$ according to the pseudo-random TDM-MIMO scheme. We define the ordered set ${\cal {P}}_{l} = \{p | {l_p} = l\}$ for the $l$th Tx-A, $l \in \{0,1,...,L_{{t}}-1\} $, and then let ${\bf{\hat x}}_{k,r,l}^{{\rm{PC}}} = {\left[ {{\bf{\hat x}}_{k,r}^{{\rm{PC}}}} \right]_{{\cal {P}}_{l}}} \in \mathbb{C}^{P/L_t}$. Similar to RE, the VE can also be formulated as a CS problem
 \begin{equation}
 \label{equ:Rve}
 	{\bf{\hat x}}_{k,r,l}^{{\rm{PC}}} = {\bf{\Phi}}_{l}  {\tilde {\bf{F}}}_{\rm VE} {\bf{z}}_{k,r,l} \text{,}
 \end{equation}
where ${\bf{\Phi}}_{l}=\left[{\bf{I}}_{P}\right]_{{\cal{P}}_{l},:}\in \mathbb{C}^{(P/L_t) \times P}$ is the measurement matrix for the $l$th Tx-A, ${\tilde {\bf{F}}}_{\rm VE} \in \mathbb{C}^{P \times \rho_{\rm VE} P}$ with $\rho_{\rm VE} \ge 1$ is the dictionary matrix consisting of the first $P$ rows of the $\left( \rho_{\rm VE}P \right)$-order DFT matrix, and ${\bf{z}}_{k,r,l} \in\mathbb{C}^{\rho_{\rm VE} P}$ is the sparse representation to be estimated.
Similarly to solving \eqref{equ:Rcs}, we apply OMP to \eqref{equ:Rve} to obtain the sparse estimate ${\hat{\bf{z}}}_{k,r,l}$ of ${{\bf{z}}}_{k,r,l}$, and the CFAR detection is subsequently used to determine the most significant element in ${\hat{\bf{z}}}_{k,r,l}$ with the index ${\hat p}_{k,r,l} \in \{0,1,...,\rho_{\rm VE}P-1\}$. Therefore, the velocity estimate $\hat{v}_{k,r,l}$ from the $l$th Tx-A and $r$th Rx-A can be obtained as $\hat{v}_{k,r,l} ={\rm{shift}}\left({\hat p}_{k,r,l}\right) \lambda /\left(2\rho_{\rm VE}PT_0 \right)$, where ${\rm{shift}}\left(x\right) \buildrel \Delta \over = x$ for $x \in \left[ 0,\rho_{\rm VE}P/2\right)$ and $x-\rho_{\rm VE}P$ for $x \in \left[ \rho_{\rm VE}P/2,\rho_{\rm VE}P\right)$.
The final estimate $\hat{v}_k$ of $v_k$ is obtained by averaging the estimates from all antenna pairs, i.e., $\hat{v}_k = \frac{1}{{{L_t}{L_r}}}\sum\nolimits_{l = 0}^{{L_t} - 1} {\sum\nolimits_{r = 0}^{{L_r} - 1} {{{\hat v}_{k,r,l}}} } $.

\subsection{Angle Estimation}
In the conventional TDM-MIMO scheme, IF signals derived from adjacent Tx-As have an additional phase shift \cite{v&r}. However, the pseudo-random TDM-MIMO scheme with CS algorithms proposed in this paper enables accurate AE without the necessity for velocity-induced PC (referred to as PC $\rm{II}$). Based on the previous estimation results, we define an AE vector ${\bf \hat a}_k \in \mathbb{C}^{L_tL_r}$ for the $k$th target as $\left[ {\bf \hat a}_k \right]_{l L_r+r} = \left[{\hat{\bf{z}}}_{k,r,l}\right]_{{\hat p}_{k,r,l}}$, where ${\hat{\bf{z}}}_{k,r,l}$ can be obtained by solving \eqref{equ:Rve}. 
To obtain the angle estimate $\hat \theta_k$ of the $k$th target from ${\bf \hat a}_k$, we construct a matrix ${\bf{\tilde{F}}}_{\rm AE} \in \mathbb{C}^{L_t L_r \times \rho_{\rm AE}L_t L_r}$ with $\rho_{\rm AE} \ge 1$, where $\left[{\bf{\tilde{F}}}_{\rm AE} \right]_{i,n_\theta} = \exp \left[ {-j \pi i \sin \frac{(n_\theta-\rho_{\rm AE}L_t L_r/2) \pi} {\rho_{\rm AE}L_t L_r} } \right]$, $i \in \left\{ 0,1,...,L_tL_r-1 \right\}$, and $n_{\theta} \in \left\{ 0,1,...,\rho_{\rm AE}L_t L_r-1 \right\}$. 
The index corresponding to the angle component of the $k$th target ${\hat n}^{\rm AE}_k$ is determined by $\hat n_k^{{\rm{AE}}} = \mathop {\arg \max }\limits_{{n_\theta }} \left| {\bf{\hat a}}_k^{\rm{T}}{\left[ {{{{\bf{\tilde F}}}_{\rm AE} }} \right]_{:,{n_\theta }}}\right|$, and the angle estimate $\hat{\theta}_k$ of $\theta_k$ can be obtained as $\hat{\theta}_k =  \frac{({\hat{n}}^{\rm AE}_k-{\rho_{\rm AE}L_t L_r}/2) \pi} {{\rho_{\rm AE}L_t L_r} } $. 
\vspace{-3mm}

\section{Data Demodulation of MmWave-LoRadar}
For communication, our focus is directed towards the data transmission process, presupposing that ideal synchronization has been accomplished, i.e., $\bar \tau + \bar v t /c=0$.
The task of communication is to determine $h_p$ from the received signal at the Com-Rx. First, the Com-Rx down-converts ${\bar r_p}\left( t \right) $ to obtain the baseband signal ${{\bar r}_p^{\rm{BB}}} \left( t \right) = {{\bar r}_p}\left( t \right)e^{-j2 \pi f_c t}$.
In LoRa demodulation, the sampling rate of analog-to-digital converters (ADCs), denoted as $\bar f$, is required to be higher than $B$ for reliable data demodulation \cite{LoRa}. This indicates that an ultra-high-speed ADC is required for mmWave-LoRadar due to the large bandwidth.
As a remedy, we consider the AIC at the Com-Rx with the compressed sampling index set ${\bar{\mathcal{M}}}=\left\{\bar{m}_{0},\bar{m}_{1},...,\bar{m}_{\bar{N}-1}\right\}$, where $0\leq \bar{m}_{0}<\bar{m}_{1}<...<\bar{m}_{\bar{N}-1}\leq \bar{N}_{\max}-1$ and $\bar{N}_{\mathrm{max}}\buildrel \Delta \over=\left\lfloor \bar{f} T\right\rfloor$. Note that $\bar N \le \bar N_{\max}$. On this basis, we obtain the received sequence with CS, denoted as ${{\bar {\bf r}}}_{{p}}^{\rm{AIC}} \in \mathbb{C}^{\bar N}$ for the $p$th symbol, which is given by
\begin{align}
\left[{{\bar {\bf r}}}_{{p}}^{\rm{AIC}} \right]_{\bar n} = {{\bar r}_p^{\rm{BB}}}\left( t \right) \left| {_{t = p{T_0} + \bar \tau  + \bar{m}_{\bar{n}}/{\bar{f}}}} \right. ,
\end{align}
where $\bar{n} \in \{0,1,...,\bar{N}_{\mathrm{max}}-1\}$.
The de-chirp sequence with CS ${{\bar {\bm r}}}_{{p}} \in \mathbb{C}^{\bar{N}}$ is obtained by mixing a compressively-sampled digital down-chirp signal ${{\bar {\bf r}}}_{{\rm down}}\in \mathbb{C}^{\bar{N}}$, i.e.,
\begin{align}
\label{equ:dechirp}
\hspace{-1.8mm} {{\bar {\bf r}}}_{{p}} & = {{\bar {\bf r}}}_{{p}}^{\rm{AIC}} \odot {{\bar {\bf r}}}_{\rm down}  \nonumber \\
& {\overset{3}{\operatorname*{\approx}} } \left\{ 
{\begin{array}{*{20}{l}}
 {{\bar \alpha _p} e^{j2\pi \left[(\frac{h_p}{T}) \frac{\bar{m}_{\bar{n}}}{ {\bar{f}}}\right]},}&{{\bar{m}_{\bar{n}}} \in \left[ 0, \bar{N}_{c} \right)} \\
{{\bar \alpha _p}{e^{j2\pi \left[(\frac{h_p}{T}-B) \frac{\bar{m}_{\bar{n}}}{ {\bar{f}}}+{\bar {\varphi_p}}\right]}},}&{\bar{m}_{\bar{n}} \in \left[ \bar{N}_{c},\bar{N}_{\max}\right)}\\
 \end{array}} \right. ,
\end{align}
where $\bar{N}_{c} = \left\lfloor \bar f \left( T-\frac{h_p}{B}\right)\right\rfloor$ , ${\bar {\varphi}}_p = e^{j\pi \frac{h_p\left(H-h_p\right)}{H}}$, and ${\left[ {{{\bar {\bf{r}} }_{{\rm{down}}}}} \right]_{\bar n}} = \exp\left\{{j\pi \left[-\frac{B}{T}\left(\frac{\bar{m}_{\bar{n}}}{\bar{f}}\right)^2+ \frac{B \bar{m}_{\bar{n}}}{ {\bar{f}}}\right]}\right\}$. 
${\overset{3}{\operatorname*{\approx}} }$ is because $\bar v t/c$ can be neglected in a single symbol.
The compressed de-chirp sequence can be formulated based on CS as
\begin{equation}
	{{\bar {\bf r}}}_{{p}} = {\bf{\bar \Phi}} {\bf{\bar F}} {{\bar {\bf x}}}_{{p}},
\end{equation}
where ${\bf{\bar \Phi}}=\left[{\bf{I}}_{\bar{N}_{\mathrm{max}}}\right]_{{\bar{\mathcal{M}}},:}\in \mathbb{C}^{\bar{N} \times \bar{N}_{\max}}$, ${\bf{\bar F}} \in \mathbb{C}^{\bar{N}_{\max}\times \bar{N}_{\max}}$ is the $\bar{N}_{\max}$-order DFT matrix, and ${{\bar {\bf x}}}_{{p}} \in \mathbb{C}^{\bar{N}_{\max}}$ is the sparse vector to be estimated. Similarly, by employing OMP, the estimate ${{\bf{\hat{\bar x}}}}_p$ of ${{\bf{{\bar x}}}}_p$ can be obtained. The demodulated value $\hat{h}_p$ of the $p$th transmit symbol can be determined by
\begin{align}
    \hat{h}_p=\underset{h\in\{0,1,\ldots,H-1\}}{\arg \max} \left|\left[{{\bf{\hat{\bar x}}}}_p\right]_h\right|+\left| \left[{{\bf{\hat{\bar x}}}}_p\right]_{h+\frac{\bar{f}-B}{B}H}\right|.		
\end{align}

\vspace{-6mm}
\section{Simulation Results}
\begin{figure*}[]
	\vspace*{-10mm}
	
	\begin{minipage}[b]{0.5\linewidth}
		\centering
		\vspace*{-1mm}
		\subfloat[]{		\includegraphics[width=5cm, keepaspectratio]{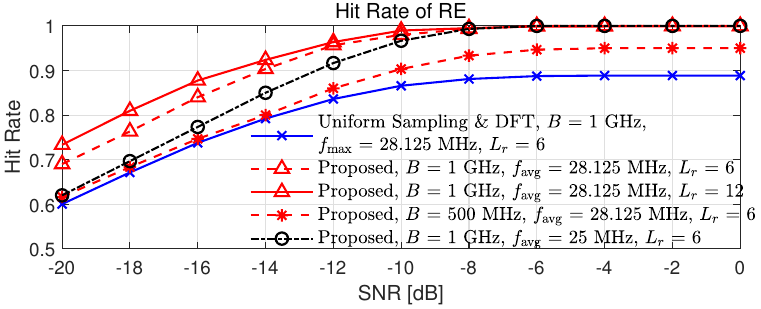}}
		\subfloat[]{		\includegraphics[width=5cm, keepaspectratio]{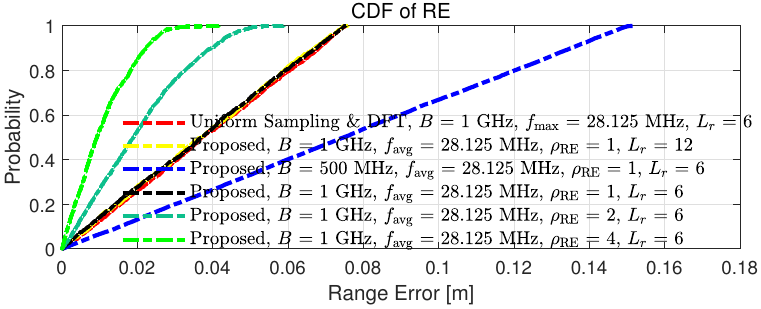}}\\
		\vspace*{-2mm}
		\subfloat[]{
			\includegraphics[width=5cm, keepaspectratio]{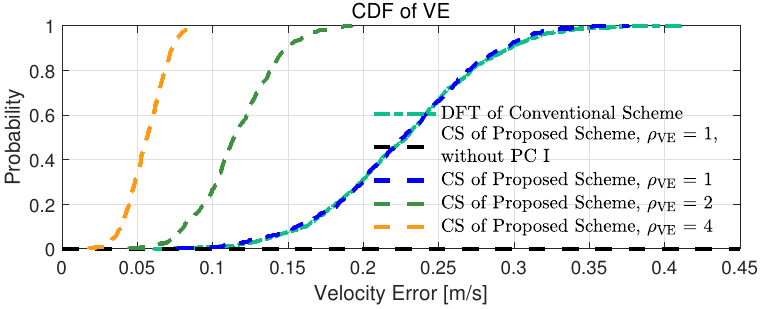}}
		\subfloat[]{
			\includegraphics[width=5cm, keepaspectratio]{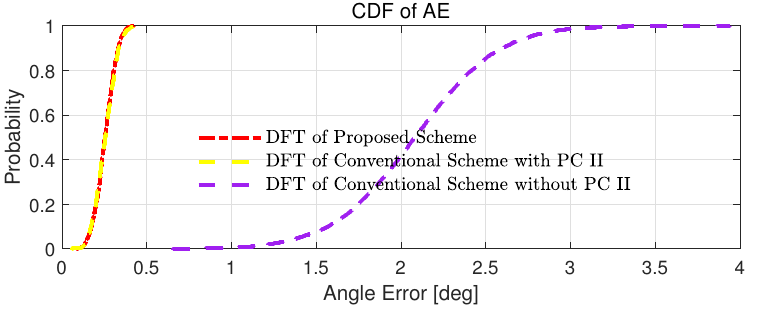}} 
	\end{minipage}
	\hspace{12mm}
	\centering
	\subfloat[]{
		\includegraphics[width=6.4cm, keepaspectratio]{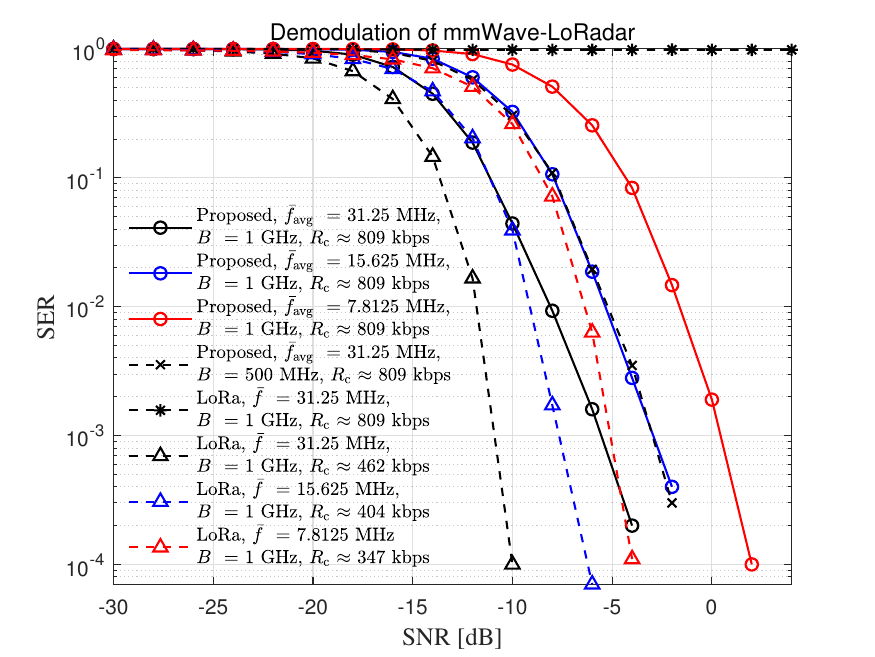}}
	\captionsetup{font={footnotesize}, name = {Fig.}, labelsep = period}
	\caption{{Simulations for mmWave-LoRadar: (a) Hit rate for RE; (b) CDF for RE; (c) CDF for VE; (d) CDF for AE; (e) SER for demodulation.} }
	\vspace*{-7mm}
\end{figure*}
To validate the proposed mmWave-LoRadar for vehicular ISAC, simulations are conducted under the following parameters: $f_{{c}} = 77$\,GHz, $B = 1$\,GHz, $N_{\rm{SF}} = 14$, $T = 16.4$\,$\mu$s, $T_{\rm GI} = T_{\rm mix} = 0.5$\,$\mu$s, $P=120$, $L_{{t}} = 2$, and $L_{{r}} = 6$. The white Gaussian noise is added to the echo signal and the communication received signal according to the signal-to-noise ratio (SNR).
For the communication channel, parameters are generated using these settings.
For the sensing model, $K=6$, $r_k$, $v_k$, and $\theta_k$ are uniformly distributed in $[0,70]$\,m, $[-50,50]$\,m/s, and $[-60^\circ,60^\circ]$, respectively.
To analyze the impact of $B$, we also select $B = 500$\,MHz with $N_{\rm{SF}}=13$ for keeping the same $T$.

For sensing, the baseline scheme is called the ``{uniform sampling \& DFT scheme}'', which can be referred to in \cite{automotive,v&r}. 
In simple terms, this scheme replaces the CS with uniform sampling and DFT.
For the hit rate of RE, we define the situation
$\left|\hat{\tau}_kc/2-r_k\right|< \frac{c\left(T-{T_{\rm{mix}}}\right)}{2BT}$ as a ``hit'', i.e., the $k$th target is successfully detected. 
{For the proposed CS scheme, the compression ratio in RE is defined as $\eta_{\rm{RE}} \buildrel \Delta \over = N/N_{\max}$, $f_{\max} = 31.25$\,MHz, $N_{\max} = 512$, and the average sampling rate for RE $f_{\rm{avg}}$ is defined as $f_{\rm{avg}} = \eta_{\rm{RE}} f_{\max}$.
To illustrate the advantages of the MMV-OMP, we provide the performance of the hit rate under different SNRs with $f_{\rm{avg}} = 28.125$\,MHz and $f_{\rm{avg}} =  25$\,MHz, respectively.
As depicted in Fig. 4(a), a lower $f_{\rm{avg}}$ indicates fewer sampling points, which leads to poorer noise resistance performance and a lower hit rate.
Similarly, for the same $f_{\rm{avg}}$, a lower $B$ also exhibits a degradation in hit rate performance, which is because of the reduction in the time-bandwidth product.
In order to ensure a fair comparison, the sampling frequency for the ``uniform sampling \& DFT scheme'' is set to $f_{\max} = 28.125$\,MHz.
In contrast, the proposed CS scheme demonstrates superior performance. 
This is because uniform sampling results in a reduction in the maximum unambiguous range $R_{\rm{max}} = \frac{c f_{\rm{max}} (T - T_{\rm{mix}})}{2B}$, therefore the plateau in high SNRs is lower.}
Additionally, as indicated by the cumulative distribution functions (CDFs) of range error $\left|\hat{\tau}_kc/2-r_k\right|$ in Fig. 4(b) at SNR = $10$ dB, the redundant dictionary results in a performance improvement. 
Under the ``uniform sampling \& DFT scheme'', the CDF in RE is close to that of CS with $\rho_{\rm{RE}}=1$, which is because both schemes are based on the same grid resolution.
{The poorer range resolution caused by the decrease in $B$ manifests on the CDF curve as a broadening of the RE error distribution.
Additionally, due to the diversity gain, increasing $L_r$ can improve the hit rate under low SNRs. However, it does not have a significant impact on the CDF.}

Fig. 4(c) depicts CDFs of velocity error $\left|\hat{v}_k-v_k\right|$ for VE at SNR = $10$ dB. It is evident that without PC $\rm{I}$, VE is not achievable, which demonstrates the necessity of the proposed PC $\rm{I}$ for mmWave-LoRadar. 
For proposed CS schemes, the velocity PC matrix ${\bf{R}}^{\rm{PC}}_{k} $ is obtained using ${\hat{\tau}}_k$ derived by solving \eqref{equ:Rcs} in Section III-B with $f_{\rm{avg}} = 28.125$\,MHz.
The conventional scheme refers to the ``uniform sampling \& DFT scheme'' which performs PC $\rm{I}$ for VE with $f_{\max} = 28.125$\,MHz.
In Fig. 4(c), an increasing $\rho_{\rm{VE}}$ leads to a better CDF, as the increased $\rho_{\rm{VE}}$ facilitates the selected supports closer to $v_k$, $\forall k$. 
Although the CDF of $\rho_{\rm{VE}} = 1$ is similar to that of the conventional scheme, the proposed CS scheme achieves a maximum unambiguous velocity $L_t$ times higher. 
Fig. 4(d) depicts CDFs of angle error $| \hat{\theta}_k-\theta_k |$ for AE at SNR = $10$ dB. 
$\rho_{\rm{AE}}$ is set to $15$ to achieve a $1^\circ$ AE accuracy. 
The proposed AE scheme demonstrates superior performance compared to the conventional scheme without PC $\rm{II}$ and slightly outperforms the scheme with PC $\rm{II}$.

For communication, $R_{\rm{c}}$ is the communication rate. 
In CS, $\bar{f}=2$\,GHz, ${\bar{\eta}} \buildrel \Delta \over =\bar{N}/\bar{N}_{\max}$ and $\bar{f}_{\rm{avg}} ={\bar{\eta}} \bar{f}$ are the compression ratio and average sampling rate for communication, respectively. 
The symbol error rate (SER) is adopted as the performance metric.
The ``LoRa scheme'' refers to demodulation through uniform sampling and DFT \cite{LoRa}.
 {In Fig. 4(e), comparing proposed CS schemes, we can observe that the lower $\bar{f}_{\rm{avg}}$ at a fixed SNR results in a higher SER, indicating that by adjusting ${\bar{\eta}}$, a data reduction in $\bar{N}$ can be achieved while maintaining acceptable communication performance.
Comparing schemes with the same number of sampling points, the SER of the ``LoRa scheme'' is better than that of the proposed CS scheme. }
However, the ``LoRa scheme'' can only correctly demodulate ($N_{\rm{SF}}+\log_{2}{{\bar{\eta}} }$) bits per symbol, resulting in a lower $R_{\rm{c}}$. 
If this scheme is used to demodulate $N_{\rm{SF}}$ bits, the SER approaches $1$, rendering the demodulation impractical. 
This demonstrates that employing the ADC for demodulation necessitates a higher $\bar{f}$ to maintain the same $R_{\rm{c}}$ compared to the proposed CS scheme.
 {For $B = 500$\,MHz and $\bar{f}_{\rm{avg}} = 31.25$\,MHz, the decrease in system performance is related to the time-bandwidth product. Compared to the proposed CS scheme with $B = 1$\,GHz and $\bar{f}_{\rm{avg}} = 31.25$\,MHz, we can find the performance drops by approximately $3$\,dB.}
Additionally, $\bar{f}_{\rm{avg}}$ is close to the IF component of sensing, i.e., $f_{\rm{avg}} = 28.125$\,MHz, which suggests the hardware sharing with AIC.
\vspace{-3mm}

\section{Conclusions}
In this paper, we introduced mmWave-LoRadar, an ISAC solution that combined LoRa modulation with FMCW in the mmWave band. This design not only preserved the communication capabilities of LoRa but also enabled high-precision radar sensing with reduced sampling rates. Our derivation revealed the presence of dual discontinuity issues in mmWave-LoRadar, highlighting the necessity for a novel signal processing framework. Based on the proposed CS-based hardware design and signal processing schemes, we addressed the challenges associated with both communication and sensing tasks. 
Through the above design, we realized waveform, spectrum, and hardware sharing, which validated the proposed ISAC system as a promising solution for mmWave vehicular networks and other domains.
The mmWave-LoRadar solution boasts outstanding scalability and can be further integrated with other existing chirp-based ISAC modulation schemes to meet the demands of data-intensive communication networks.

\end{document}